\documentclass[ twocolumn,superscriptaddress,10pt, amsmath,amsfonts,amssymb,preprintnumbers]{revtex4}
\pdfoutput=1
\usepackage[T1]{fontenc}\usepackage[latin1]{inputenc}
\usepackage{dcolumn,graphicx,color,booktabs,microtype}
\usepackage[charter]{mathdesign}

\renewcommand{\figurename}{Fig.}
\renewcommand{\tablename}{Table}
\makeatletter\renewcommand{\fnum@figure}[1]{\figurename~\thefigure.}\makeatother
\makeatletter\renewcommand{\fnum@table}[1]{\tablename~\thetable.}\makeatother
\usepackage[colorlinks,plainpages=false,linkcolor=black,urlcolor=blue,citecolor=black,pdfpagemode=UseNone,pdfstartview=FitH]{hyperref}

\begin{document}

\title{Fermi surface of Ba$_{1-x}$K$_x$Fe$_2$As$_2$ as probed by angle-resolved photoemission. }

\author{V.~B.~Zabolotnyy}
\author{D.~V.~Evtushinsky}
\affiliation{Institute for Solid State Research, IFW-Dresden, P.O.Box 270116, D-01171
Dresden, Germany}

\author{A.~A.~Kordyuk}
\affiliation{Institute for Solid State Research, IFW-Dresden, P.O.Box 270116, D-01171 Dresden, Germany}
\affiliation{Institute of Metal Physics of National Academy of Sciences of Ukraine, 03142 Kyiv,  Ukraine}

\author{D.~S.~Inosov}
\affiliation{Max-Planck-Institute for Solid State Research, Heisenbergstraße 1, D-70569
Stuttgart, Germany}

\author{A.~Koitzsch}
\affiliation{Institute for Solid State Research, IFW-Dresden, P.O.Box 270116, D-01171 Dresden,
Germany}

\author{A.~V.~Boris}
\affiliation{Max-Planck-Institute for Solid State Research, Heisenbergstraße 1, D-70569
Stuttgart, Germany}
\affiliation{Department of Physics, Loughborough University,
Loughborough, LE11 3TU, United Kingdom}

\author{G.~L.~Sun}
\author{C.~T.~Lin}
\affiliation{Max-Planck-Institute for Solid State Research, Heisenbergstraße 1, D-70569 Stuttgart,
Germany}

\author{M.~Knupfer}
\author{B.~B\"{u}chner}
\affiliation{Institute for Solid State Research, IFW-Dresden, P.O.Box 270116, D-01171 Dresden,
Germany}
\author{A.~Varykhalov}
\author{R.~Follath}
\address{Elektronenspeicherring BESSY II, Helmholtz-Zentrum Berlin für Materialien und Energie,\\ Albert-Einstein-Strasse 15, 12489 Berlin, Germany}
\author{S.~V.~Borisenko}
\affiliation{Institute for Solid State Research, IFW-Dresden, P.O.Box 270116, D-01171 Dresden,
 Germany}

\begin{abstract}
Here we apply high resolution angle-resolved photoemission spectroscopy (ARPES) using a
wide excitation energy range to probe the electronic structure and the Fermi surface
topology of the Ba$_{1-x}$K$_x$Fe$_2$As$_2$ ($T_\textup{C}$ = 32\,K) superconductor. We
find significant deviations in the low energy band structure from that predicted in
calculations. A set of Fermi surface sheets with unexpected topology is detected at the
Brillouin zone boundary. At the X-symmetry point the Fermi surface is formed by a shallow
electron-like pocket surrounded by four hole-like pockets elongated in $\Gamma-$X and
$\Gamma-$Y directions.
\end{abstract}

\maketitle

\noindent Since the discovery of superconductivity in Fe-based pnictides \cite{Kamihara2006, Kamihara2008}
with critical temperatures comparable to those of the famous layered cuprate
high-$T_{\textup{c}}$ superconductors \cite{Takahashi, Kito, Chen_Nature, ChenPRL,
rotter107006, Wu}, the interest raised by these materials seem to excel the one evoked by
the discovery of cuprates.  Important conclusions as for the nature of the
superconductivity \cite{Richard}, symmetry of the superconducting order parameter
\cite{Sknepnek, Parish, ChenBCS} and its absolute value \cite{Zhao, Wray, KondoPRL101}
are currently being made, while the electronic structure of the novel type of
superconductors remains controversial.
 According to early angle-resolved photoemission studies (ARPES) \cite{LiuPRL101, Yang,
Ding, Zhao}, the band structure calculations \cite{LiuPRL101, Ma, Singh0, Nekrasov,
Singh, Mazin} seem to capture the most essential ingredients of the low-lying electronic
structure. The basic result of the band structure calculations is that the inequivalence
of the As sites in the ``parent'' compound BaFe$_2$As$_2$ (BFA) results in a Fermi
surface (FS) folding into two concentric hole-like FS sheets centered at the
$\Gamma$-point and double-walled electron-like pockets at the X and Y points. In another
\emph{parent} Fe-based system LaOFeP even more remarkable agreement was found\cite{Shen}.
However, in the case of \emph{doped} superconducting BKFA the experimental reports
concerning the FS topology  are far from been equivocal.
  According to Ref. \onlinecite{Ding}, there is a  single electron-like pocket covering 3\% of the BZ, whereas Zhao et al. \cite{Zhao} report two
intensity spots in the vicinity of the X/Y-point with no evidence for the electron pocket in the
normal state.

In this work, we address the issue of the low energy band structure ($E \sim -300..0$ meV)
of the Ba$_{1-x}$K$_x$Fe$_2$As$_2$ superconductor and show that there are notable
deviations in the FS topology as compared to the currently assumed band structure based
on the available calculations. Our finding affects both the theoretical works generally
aimed at reproducing the electronic structure, as well as those concentrated on a more
subtle issue of superconductivity origin in the new type of superconductors. Since the FS
topology also determines the loci in the reciprocal space where the superconducting gaps
are to be evaluated,  the correct FS topology may result in a different interpretation of
current experimental data on the superconducting gaps in the novel superconductor
\cite{DaniilGaps, Ding, Zhao, Wray, KondoPRL101}. In view of the importance of magnetic
excitations in the cuprate superconductors and extended region in the phase diagram
\cite{Chen, Alireza} of pnictide superconductors  characterized by a spin density wave
order our observations may attract attention of the scientific community to a possible
common feature between the two types of unconventional superconductors.

\begin{figure*}
\begin{center}
\includegraphics[width=0.75\textwidth]{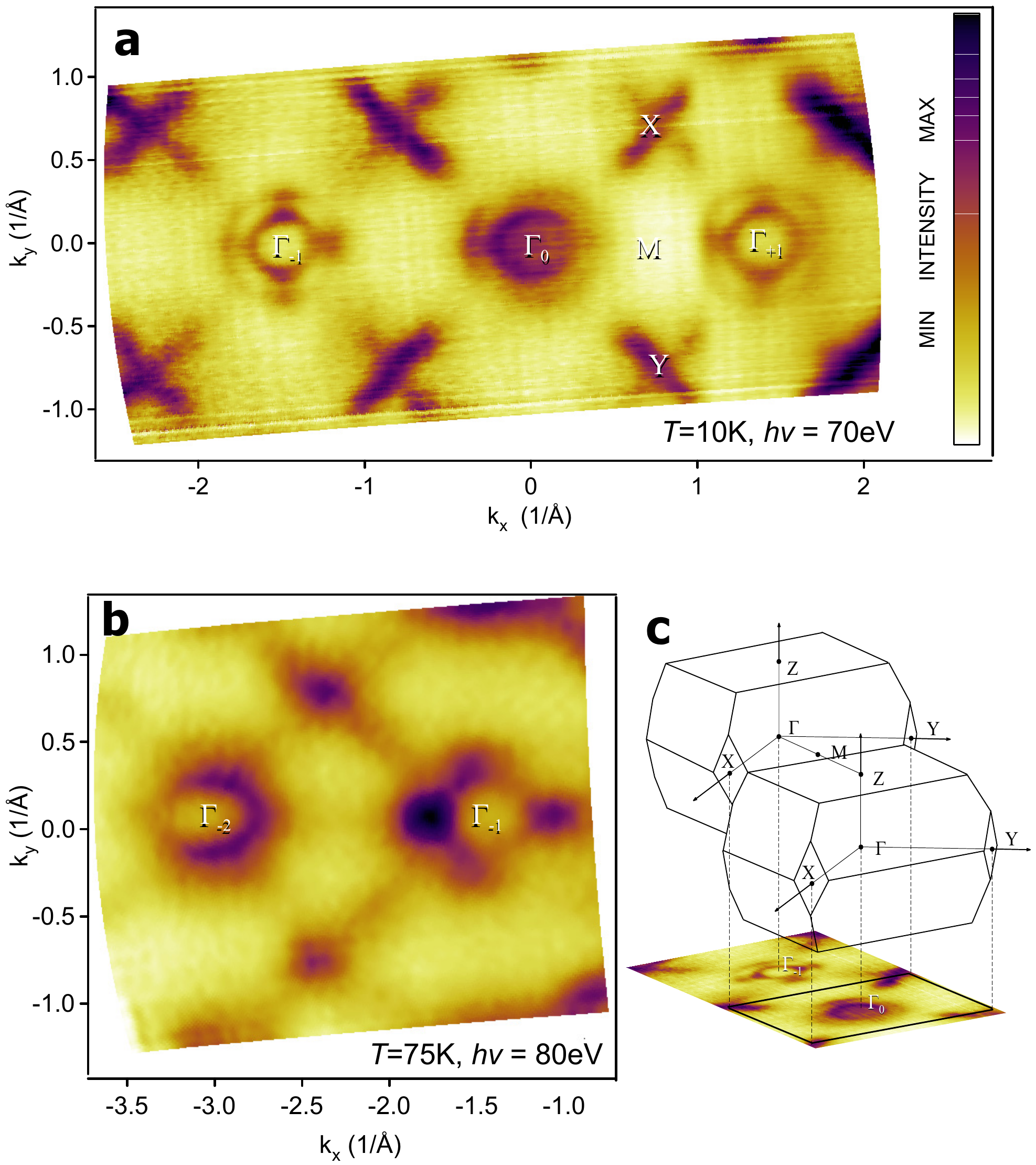}\\
\caption{\label{Images} (a, b), Momentum dependence of the photoemission intensity at
constant energy cuts for Ba$_{1-x}$K$_x$Fe$_2$As$_2$, at $T$ = 10 and 75K. (c) 3D
Brillouin zone and notation of high symmetry points. When referring to ARPES data we use
symbols $\Gamma$, X, Y and M as determinants for $k_x$ and $k_y$ values, not necessary
implying the same values of $k_z$ as in 3D case. }
\end{center}
\end{figure*}

ARPES data presented in this work were collected using synchrotron radiation
(``$1^3$-ARPES'' end station at BESSY) within the range of photon energies (20\,--\,90
eV) and various polarisations from cleaved surfaces of high quality single crystals. The
overall energy and momentum resolutions were $\sim5$\,meV  and $\sim$0.013\AA$^{-1}$
respectively for the low temperature measurements. Single crystals of
Ba$_{1-x}$K$_x$Fe$_2$As$_2$ were grown using Sn as flux in a zirconia crucible sealed in
a quartz ampoule filled with Ar. The mixtures of Ba, K, Fe, As and Sn in wt\% ratio of
(Ba$_{1-x}$K$_x$Fe$_2$As$_2$):Sn = 1:85 were then heated in a box furnace up to
850$^\circ$C and kept constant for 2 -- 4 hours to soak the sample in a homogeneous melt.
For growth the cooling rate of 3$^\circ$C/h was applied to decrease the temperature to
550$^\circ$C and then the grown crystals decanted from the flux.
\begin{figure*}
\begin{center}
\includegraphics[width=0.75\textwidth]{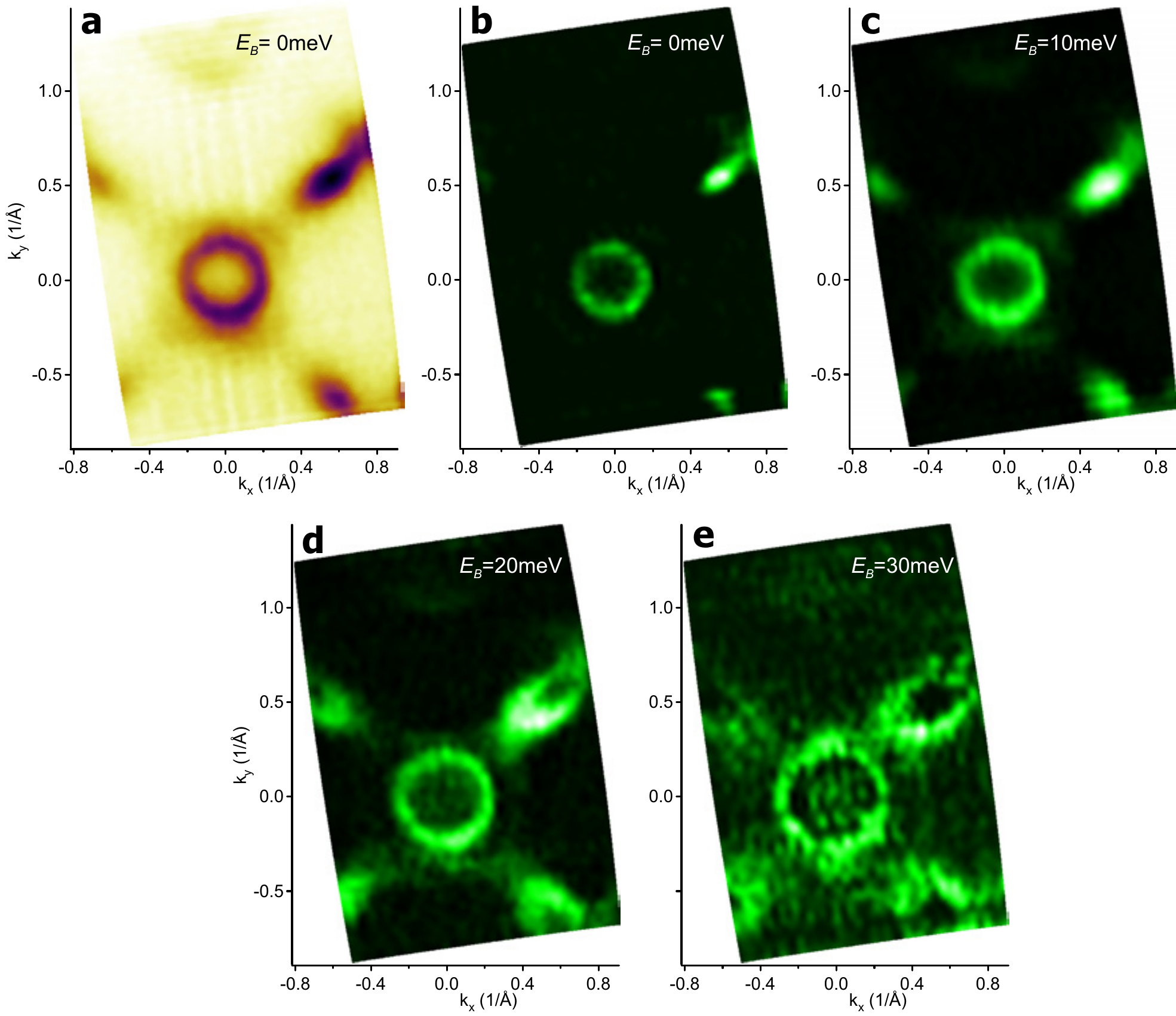}\\
\caption{\label{Images} (a) Fermi surface map of Ba$_{1-x}$K$_x$Fe$_2$As$_2$, $T$=15K,
$h\nu$=70eV. (b--e) Second derivative of photoemission intensity with respect of energy
as a function of $k_x$ and $k_y$ for a set of fixed energy cuts. The corresponding
binding energies of the cuts are give in each panel. }
\end{center}
\end{figure*}

In Fig.\,1a we show experimental FS map of BKFA measured in the superconducting state at
$T=$10\,K, which represents the photoemission intensity integrated in a small energy
window $E=E_\textup{F} \pm10$\,meV around the Fermi level (FL). Two concentric
$\Gamma$-centered FS sheets are observed in agreement with previous studies \cite{Zhao,
Ding}. Note a significant difference in the intensity distribution at different $\Gamma$
points. The intensity pattern apparently repeats itself every second $\Gamma$ point. This
periodicity is not surprising, since such an effect is expected for a hypothetical
``parent'' compound with equivalent As sites. Additionally, the 3D structure of the
Brillouin zone (BZ) also leads to a similar effect. In the simplest model of the
photoemission the intensity distribution like the one displayed in Fig.\,1a may be
thought of as  a cut through the 3D reciprocal space by the Ewald's sphere projected on the
$k_x\,k_y$ plane \cite{Cloetta} (see panel c). At sufficiently large excitation energy,
i.e. large radius of the Ewald's sphere, every second point with $k_{x,y}=0$ would have
the same $k_z$, while for the neighboring points their $k_z$ values would sum up to
$\pi/c$. This, in turn, implies the periodic change of the initial state symmetry and
thus photoemission intensity, when moving to a neighboring BZ. It is important to stress
that these intensity variations do not necessarily infer large $k_z$ dispersion. On the
contrary, as compared to the strong variation in the intensity distribution, the size of
the $\Gamma$ barrels centered at $\Gamma_0$ and $\Gamma_{\pm1}$ points turn out to be
practically the same, which implies small dependence of $\varepsilon(k_x, k_y, k_z)$ on
$k_z$.  Albeit an exception \cite{Nekrasov} this is in sharp contrast with the results of
most of the currently available band structure calculations \cite{Ma, LiuPRL101, Singh0},
though inline with the anisotropy of the transport properties \cite{Wang}. Further
differences from the band structure calculations can be derived from the shape of the
inner $\Gamma$-centered FS sheet. While in the experimental data the corners of this
square-like FS sheet are directed towards the next $\Gamma$ point, in the theoretical
calculations the picture is rotated by 45$^\circ$.

    However, the major discrepancy between experiment and \textit{ab initio} calculations
 is observed near the X/Y-point. According to the
calculations, one expects a sizeable double-walled electron pocket. Instead, as it can be
seen in Fig. 1, there is a propeller-like structure consisting of five small FS sheets
near the X/Y-point: a central patch situated exactly at X point and four blades extended
along $\Gamma$-X directions surrounding the central pocket. In Fig. 1b we show a similar
FS map measured at $T$=75K. A detectable weakening and redistribution of intensity in the
propeller-like structure can be detected above this temperature, though we find that a
residual spectral weight concentrated along the `blades' persists up to room
temperature. Such temperature dependence of the intensity cannot be simply explained by
the temperature broadening effects, and implies that there must be some variation in the
electronic band structure. This is especially interesting in a view of the recent $\mu$SR
and neutron scattering experiments\cite{Park} that show disappearance of commensurate
magnetism above 70\,K, implying that the new structure may result from folding caused by
magnetic order \cite{BKFANature}.

\begin{figure*}
\begin{center}
\includegraphics[width=0.6\textwidth]{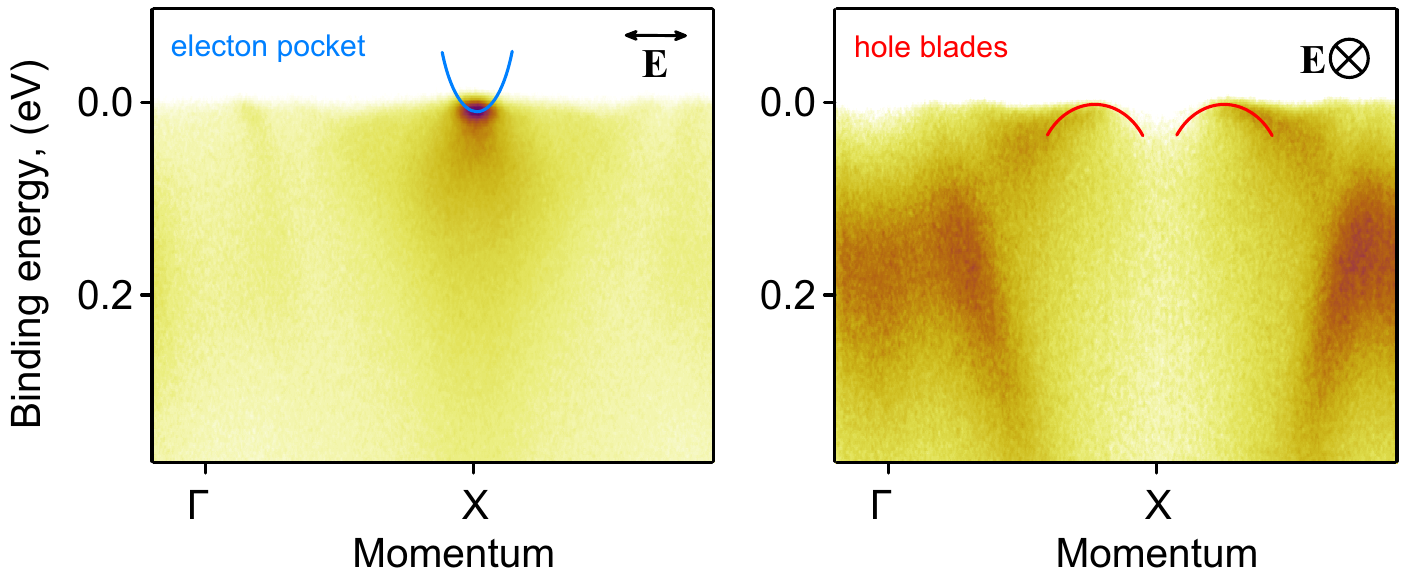}\\
\caption{\label{Images} Energy--momentum cut measured with different light polarization
along the $\Gamma$-X-$\Gamma$ direction.  (a) Polarization vector of light is parallel to
the analyzer slit. (b) Polarization vector perpendicular to the analyzer slit.  The blue
and red curves are guide to the eye corresponding to the electron pocket and blades
respectively. }
\end{center}
\end{figure*}

To study the new structure in more detail, in Fig. 2a we present a finer FS map covering
the nearest $\Gamma$ and X points.
To better display the dispersion of relatively broad bands, in the remaining panels
(b--e) we plot the second derivative of the photoemission intensity with respect to
energy as a function of quasimomentum $(k_x,k_y)$. As can be seen the size of the blade
pockets clearly increases similar to the size of the $\Gamma$ barrel, signalling
hole-like character of the both. At the same time the feature centered exactly at the X
point in the cut made at $E$=0\,meV seem to disappear, which means that its intensity must
be caused by the very bottom of the electron-like pocket. This conclusion can also be
supported by separate  energy--momentum cuts made along the $\Gamma$-X-$\Gamma$
direction as shown in Fig. 3. Depending on the light polarization one may separately
enhance the photoemission either from the hole-like blades or from the electron pocket.
This means that the X pocket and blades belong to different bands, since it would be
unlikely for a single band to change its symmetry, i.e. its orbital character, at such a
small distance in momentum space that separates the blades from the central pocket. While
in the panel (b) one can find two hole bands that support the blade FS pockets situated
symmetrically around X point and dispersing down below the FL, in the panel (a) the
spectral weight from the X pocket is mainly concentrated in the close vicinity of the FL,
which is the expected picture for a shallow electron-like band been cut by the Fermi
function.

As we can now summarize, the currently available band structure calculations
\cite{LiuPRL101, Ma, Singh0} do not reproduce  the experimentally observed electronic
structure exactly. Besides the issue of the $k_z$ dispersion and form of the inner
$\Gamma$-centered FS sheet, the deviations detected in the vicinity of the X/Y points are
the most salient. The importance of the observed topology near the X/Y-point is that it
significantly influences the density of states near the FL, which can be crucial for
superconductivity \cite{Cappelluti, Radtke}.
In conclusion we want to stress that the precise knowledge of the FS allows one to understand the
origin of many physical properties of a solid, like the  kinetic coefficients \cite{Evtushinsky},
and the electronic susceptibility \cite{Inosov}. Details of the electronic structure determine the
propensity of a system to additional ordering \cite{PRLTaSe2}. In the case of superconductors, both
calculations of critical temperature, as well as experimental determinations of the superconducting
gap, set extremely strict requirements for the input band structure data. If doping and pressure
\cite{Alireza, Chen} are so important for the superconductivity and influence first of all the
shape of the FS, the precise knowledge of the low energy electronic structure will play a vital
role in understanding superconductivity in the novel materials.

The project was supported, in part, by the DFG under Grant No. KN393/4, KN393/12 and BO 1912/2-1. We
are grateful to I. Eremin,  O. K. Andersen,  L. Boeri, I. Mazin and M. Rümelli for fruitful
discussions.

\end{document}